\documentstyle[preprint,aps,epsf]{revtex}
\def\bea{\begin{eqnarray}}
\def\eea{\end{eqnarray}}
\def\beaz{\begin{eqnarray*}}
\def\eeaz{\end{eqnarray*}}
\def\bsea{\begin{subeqnarray}}
\def\esea{\end{subeqnarray}}
\def\bsea{\begin{eqnarray}}
\def\esea{\end{eqnarray}}
\def\slabel{\label}
\def\be{\begin{equation}}
\def\ee{\end{equation}}
\def\bez{\begin{equation*}}
\def\eez{\end{equation*}}
\def\nonu{\nonumber}
\def\[{\left[}
\def\]{\right]}
\def\({\left(}
\def\){\right)}

\def\intdke{\int \limits_E {\rm d}^3 k \,}
\def\intdkm{\int \limits_M {\rm d}^3 k \,}
\makeatletter
 \def\Tr{\mathop{\operator@font Tr}\nolimits}
\makeatother
\makeatletter
 \def\max{\mathop{\operator@font max}\nolimits}
\makeatother
\def\gm{\gamma^{\mu}}
\def\gn{\gamma^{\nu}}

\def\gl{\gamma^{\lambda}}
\def\gr{\gamma^{\rho}}

\def\Gn{\Gamma^{\nu}}
\def\Gm{\Gamma^{\mu}}

\def\pslash{\hat p}

\def\kslash{\hat k}

\begin{document}

\tighten
\draft
\preprint{
\parbox[t]{30mm}{{UG-10/95}\\
\phantom{UG-10/95}\\
\phantom{UG-10/95}\\
\phantom{UG-10/95}\\
\phantom{UG-10/95}
}}

\title{
\boldmath\mbox{$(2+1)$}-Dimensional
\boldmath\mbox{$\rm QED$} with 
Dynamically \\ Massive Fermions  
in the Vacuum Polarization}

\author{V.P.~Gusynin$^{1,2}$, A.H.~Hams$^1$ and M.~Reenders$^1$}
\address{$^1$
Institute for Theoretical Physics,\\
University of Groningen,
9747 AG Groningen, The Netherlands}
\address{$^2$
Bogolyubov Institute for Theoretical Physics,\\
 National Academy of Sciences of Ukraine, 
252143 Kiev, Ukraine}

\date{September 19, 1995}

\maketitle

\begin{abstract}
We study chiral symmetry breaking in $\rm QED_3$ 
with $N_f$ flavors of four-component fermions. 
A closed system of 
Schwinger-Dyson equations for fermion 
and photon propagators and the full fermion-photon 
vertex 
is proposed, which is  consistent with the Ward-Takahashi identity.
A simplified version of that set of equations is reduced 
(in nonlocal gauge) to the equation for 
a dynamical fermion mass function, where the one-loop 
vacuum polarization with dynamically massive fermions 
has been taken into account. The linearized equation 
for the fermion mass function is analyzed in real space.
The analytical solution is compared with the results of 
numerical calculations of the nonlinear integral equation in 
momentum space.
\end{abstract}


\newpage

\section{Introduction}

Quantum electrodynamics in  one temporal, and  two spatial dimensions  
with $N_f$ flavors of four-component Dirac fermions (${\rm QED_3}$) 
continues to
attract attention as a useful field-theoretical model for
studying such phenomena as confinement and chiral symmetry
breaking ($\chi\rm{ SB}$) which are out of the realm of
perturbation theory. The model has properties reminiscent of QCD and other
four-dimensional gauge theories. Thus it has an intrinsic dimensionful 
parameter $e^2$,
the coupling constant, that plays a role similar to the QCD scale $\Lambda$,
and the effective coupling $\alpha(q)$ approaches zero at large momenta
$q$ \cite{appel1}. Studying Schwinger-Dyson equation (SDE) for the fermion
self-energy under the bare vertex approximation in the Landau gauge,
Appelquist, Nash and Wijewardhana \cite{appel2} have shown the existence of a
finite critical number of fermions, $N_c$, below which the chiral
symmetry is broken and fermions acquire a dynamical mass. In the vicinity
of $N_c$ this mass is exponentially small with respect to the natural
dimensionful parameter $e^2$, showing how a hierarchy of
scales can occur in gauge theories.
These results were confirmed by Dagotto, Kogut, Koci\'c \cite{DKK1,DKK} 
in their 
Monte-Carlo simulations of non-compact lattice $\rm QED_3$.

Continuum studies of $\chi SB$ use SDE which, being an infinite chain of
equations, must be truncated in some way. The bare vertex approximation 
was criticised \cite{PenWeb,Atk} as not being consistent with the 
Ward-Takahashi (WT) identity. An effective way to truncate
SDE is to make an ansatz for the vertex satisfying certain criteria
\cite{Penn} (for a review of SDE in $\rm QED_3$ see \cite{Roberts}). 
While at the present time the latter approach seems to be the most
effective way to satisfy WT identity under truncation, it does not shed much
light on an approximation from the physical point of view. Ideally,
one should solve SDE for the vertex itself, 
but that is quite a formidable task. 
However, in the second section we propose one possible truncation scheme
consistent with WT identity not mentioned before in
the literature.

Another important point in studying the $N_f$ dependence  of $\chi {\rm SB}$ 
is that it is necessary to insert vacuum polarization effects into the 
equation for the fermion dynamical mass (in quenched approximation, chiral
symmetry is broken for any $N_f$ \cite{Hosh,Kondo,DKK}). 
In $1/N_f$ expansion the
one-loop vacuum polarization with massless fermions is used. However, as was
mentioned in \cite{Burden,Maris1}, the inclusion of
massive fermions changes drastically the infrared behavior of the model,
leading to the appearance of a confining (logarithmic) term in the 
potential at large distances. To take into account this confining nature
of the model one can study the coupled system of SDE for the fermion 
self-energy, the photon polarization tensor and the vertex (for example,
to study the coupled system of equations for them mentioned above). In 
this paper we chose another approach, replacing the fermion dynamical mass
function in the polarization tensor by a constant dynamical mass of a
fermion. While keeping the confinement property of the model this makes 
analysis much simpler. Because this approximation just corresponds to
using the bare vertex, in order to overcome the inconsistency with the WT
identity (or, at least, minimize its violation) we selected a nonlocal gauge
\cite{Simmons,Kugo} to keep the fermion wave function 
$Z(p)\equiv1$.

In the third section we reconsider the analysis of linearized SDE for a
fermion mass function, converting it into a Schr\"odinger-like equation in
real Euclidean space with an effective potential. For the vacuum polarization
with massless fermions this potential behaves as $-{\lambda/r^2}$ at large
$r$ ($\lambda=32/{3\pi^2N_f}$). From the theory of the Schr\"odinger equation
for such kinds of potentials, it is known that it has an infinite number of
bound states if $\lambda>1/4\,(N_f<N_c=128/{3\pi^2})$. However a finite 
number of bound states might exist for $\lambda<1/4$ as well. We prove that
for the potential under consideration there are no bound states for 
$\lambda<1/4$, which means that the critical number $N_c$ does indeed 
correspond to the appearance of an infinite number of bound states. 

In the case of massive fermions, the vacuum polarization leads to a Coulomb
tail in an effective potential at large $r$, however, as we show in
Section 4, this does not alter the exponential-type behavior of a
dynamical mass near the critical $N_c$. This is, of course, in agreement
with bifurcation analysis (see, for example, \cite{AtkJohnStam,KondoJMP}). 
The main difference between a vacuum polarization with massive or with 
massless fermions lies in an overall scaling factor, which is larger for 
massive fermion loops in vacuum polarization. From the physical point of
view it means that chiral symmetry breaking takes place essentially at
intermediate 
distances $1/\alpha<r<1/m$ ($\alpha=e^2N_f/8$) and is not influenced
much by large distances, where the confinement property of the model exhibits
itself. 
This situation is similar to that in ${\rm QCD_4}$ \cite{MirGusSit,Haym},
where the chiral symmetry breaking scale is also different from the scale of
confinement.

The real space approach allows us to get a good analytical solution, fitting
nicely computer calculations when $N_f$ is close to $N_c$.
In Section 5 we return to momentum space to analyze nonlinear SDE for a
fermion dynamical mass function. We give the results of computer
calculations of the dynamical mass function, which show very convincingly
how a new energy scale (mass of a fermion) appears  that is 
much smaller than the natural scale $(e^2)$ of the model. This might be 
important for a better understanding of a hierarchy of scales in unified
theories.

\section{Truncation of Schwinger-Dyson Equations}
 
In $\rm QED_3$ the Schwinger-Dyson equation for the fermion propagator 
in Min\-kow\-ski space 
is given by
\bea
S^{-1}(p)=\hat p-\frac{ie^2}{(2\pi)^3}
\intdkm \gm S(k+p)\Gn(k+p,p) D_{\mu\nu}(k),\label{dsef}
\eea
and the SDE for the photon propagator is
\bea
D^{-1}_{\mu\nu}(q)={D_b}^{-1}_{\mu\nu}(q)+\Pi_{\mu\nu}(q),\label{dseph}
\eea
where 
\bea
\Pi^{\mu\nu}(q)=\frac{ie^2}{(2\pi)^3}\intdkm {\rm Tr}
\[\gm S(k)\Gn(k,k-q)S(k-q)\].
\label{dsevp}
\eea
In Eqs.(\ref {dsef}) and (\ref {dsevp}) $\Gn(k,p)$ is the dressed
fermion-photon vertex which should satisfy the 
Ward-Takahashi identity
\bea
(k-p)_\mu\Gm(k,p)=S^{-1}(k)-S^{-1}(p)
\label{wti}
\eea
Eqs.(\ref{dsef}) and (\ref{dsevp}) for the full fermion and photon
propagators should be solved together with the equation for the full
vertex function $\Gm(k,p)$
\bea
\Gm_{ab}(k,p)&=&\gm_{ab}+\frac{ie^2}{(2\pi)^3}
\int\limits_M{\rm d}^3q\,
\[S(k+q)\Gm(k+q,p+q)S(p+q)\]_{dc}\nonu\\
&&\qquad\quad\times\, K_{cd,ba}(p+q,k+q,q),
\label{eqg}
\eea
which in its turn contains an unknown kernel $K(p,k,q)$
(fermion-antifermion scattering amplitude), and so on. Because the
system of SDE is in fact an infinite chain of integral equations for
$n$-point Green's functions, we are forced to use some truncation scheme, 
which we require to be
consistent with WT identity. One can verify that if we take the bare vertex
approximation ($\Gm=\gm)$ in Eq.(\ref{dsef}) and use the ladder 
approximation for the kernel 
$K(p,k,q)$:
\bea
K(p+q,k+q,q)_{cd,ba}
=\gamma^\lambda_{ad}\gamma^\rho_{cb}D_{\lambda\rho}(q),
\label{kern}
\eea
we get a closed system of SDE consistent with WT identity (\ref{wti})
(Eqs.(\ref{dseph}),(\ref{dsevp}) remain unaltered), see 
Fig.~\ref{fig1}. 
Indeed, let us multiply
Eq.(\ref{eqg}) with the kernel (\ref{kern}) by $(k-p)_\mu$, then taking
into account Eq.(\ref{dsef}) (with $\Gn=\gn$) we write
\bea
&&\hat k-\hat p=S^{-1}(k)-S^{-1}(p)\nonu\\
&&\hspace{-11mm}
+\frac{ie^2}{(2\pi)^3}\int\limits_M{\rm d}^3 q\,\gm S(q+k)\gn
D_{\mu\nu}(q)
-\frac{ie^2}{(2\pi)^3}\int\limits_M{\rm d}^3 q\,\gm S(q+p)\gn
D_{\mu\nu}(q),
\eea
which allows us to reduce Eq.(\ref{eqg}) to the following:
\bea
&&(k-p)_{\mu}\Gm(k,p)-S^{-1}(k)+S^{-1}(p)=
\frac{ie^2}{(2\pi)^2}\int\limits_M{\rm d}^3q\,
\gl S(k+q)\nonu\\
&&\times\,
\[(k-p)_{\mu}
\Gm(k+q,p+q)-S^{-1}(k+q)+S^{-1}(p+q)\]\nonu\\
&&\times\,
S(p+q)\gr D_{\lambda\rho}(q).
\label{ident}
\eea 

It is evident now that Eq.(\ref{ident}) has a solution satisfying WT
identity Eq.(\ref{wti}).

Due to the Ward-Takahashi identity, the vacuum polarization tensor 
Eq.(\ref{dsevp}) has the form
\bea
\Pi_{\mu\nu}(q)=(-g_{\mu\nu}q^2+q_{\mu}q_{\nu})\Pi(q^2),\label{wardvac}
\eea
where $\Pi(q^2)$ is the vacuum polarization (we recall 
that we use four-component spinors).

The general form for the full fermion propagator can be expressed as
\bea
S(p)= \frac{Z(p^2)}{\pslash-M(p^2)},
\eea
where $Z$ is the fermion wave function and $M$ is the mass function. 
Then the equation
 for the mass function is 
\bea
\frac{M(p^2)}{Z(p^2)}=\frac{ie^2}{(2\pi)^3} \intdkm \gm\gn D_{\mu\nu}(k-p)
\frac{M(k^2) Z(k^2)}{k^2-M^2(k^2)},\label{mpdse}
\eea
and for the fermion wave function
\bea
\pslash\(1-\frac{1}{Z(p^2)}\)
=\frac{ie^2}{(2\pi)^3} \intdkm \gm\kslash\gn D_{\mu\nu}(k-p)
\frac{Z(k^2)}{k^2-M^2(k^2)}.\label{zpdse}
\eea
The general tensor structure of the vertex contains eight scalar
functions \cite{Penn}; thus, together with the functions $Z(p^2)$, $M(p^2)$
and $\Pi(p^2)$,  we can write down a coupled system of integral equations 
for eleven unknown scalar functions. Perhaps, this is the simplest 
truncated set of SDE consistent with the WT identity. 
We emphasize that such a truncation scheme can be used also in four-dimensional
${\rm QED}$ with one (though important) difference: due to the superficial 
linear
divergence of the electron self-energy we should take a photon momentum as
an integration variable in order to avoid shifting variables. Certainly, 
this scheme does not satisfy the requirement of multiplicative
renormalizability, but that property is a feature of the whole theory 
rather than 
of any approximation scheme. 

Now, it is well known that $Z$ is a gauge-dependent function, and we can
use the freedom of choosing a convenient gauge to make the function
$Z(p^2)=1$. To achieve this we use a nonlocal gauge fixing procedure 
\cite{Simmons,Kugo}.
The photon propagator in a general nonlocal covariant gauge 
is defined as \cite{Landau2}
\bea
D_{\mu\nu}(q)=\(-g_{\mu\nu}+\frac{q_{\mu}q_{\nu}}{q^2}\)
\frac{1}{q^2}\frac{1}{1+\Pi(q^2)}-\xi(q^2)\frac{q_{\mu}q_{\nu}}{q^4},
\label{genph}
\eea
with $\xi(q^2)$ being a function of momentum $q$ rather than a constant.
Simmons proved \cite{Simmons} that a suitable form 
for $\xi(q^2)$ is the following:
\bea
\xi(q^2)=\frac{2}{1+\Pi(q^2)}
-\frac{2}{q^2}\int \limits_0^{q^2} \frac{{\rm d} v}{1+\Pi(v)}.
\label{gaugfie}
\eea
With this gauge function $\xi(q^2)$, the right-hand side of Eq.(\ref{zpdse}) 
vanishes when averaged over the direction of \mbox{\boldmath $k$}. 

For Eq.(\ref{mpdse}) we obtain, in Euclidean formulation,
\bea
M(p^2)=\frac{e^2}{(2\pi)^3} \intdke 
\frac{M(k^2)}{k^2+M^2(k^2)} \frac{1}{q^2}
\[\frac{2}{1+\Pi(q^2)}+\xi(q^2)\],\label{meqeucl}
\eea 
where $q^{\mu}=k^{\mu}-p^{\mu}$.
 
The system of Eqs.(\ref{dseph}), (\ref{dsevp}), (\ref{eqg}), (\ref{kern}) 
and (\ref{meqeucl}) is 
still far too complicated for an analytical study, and we postpone the 
full investigation of it to the future. Here we note only that the mass 
function $M(p^2)$ is connected to scalar functions from the vertex through
the vacuum polarization $\Pi(q^2)$. If we assume that this connection is
not crucial, we can take the one-loop approximation for the vacuum 
polarization, replacing the vertex $\Gn$ by the bare vertex $\gn$, 
and the running
mass function $M(p^2)$ by its value $M(0)$. This approximation,
though keeping the transversality condition for $\Pi_{\mu\nu}(q)$, leads to 
violation of the WT identity Eq.(\ref{wti}), however, in the nonlocal gauge
the WT identity is still approximately satisfied. 
 
Thus the vacuum polarization in the one-loop approximation takes the form
(with Euclidean momentum $q$):
\bea
\Pi(q^2)=\frac{N_f e^2}{4\pi q^2}\[2 m
+ \frac{q^2-4m^2}{q}\arctan\(\frac{q}{2m}\)\],\label{vac1loop}
\eea
where in our case we take $m\equiv M(p^2=0)$.
Taking massless fermion loops in vacuum polarization, 
we find
\bea
\Pi(q^2)=\frac{\alpha}{q},\qquad \alpha=\frac{N_f e^2}{8}.\label{vpmless}
\eea

\section{Coordinate Space Formulation}

First we study Eq.(\ref{meqeucl}) in the linearized form, i.e.
\bea
M(p^2)=\frac{e^2}{(2\pi)^3} \intdke 
\frac{M(k^2)}{k^2+m^2} \frac{1}{q^2}
\[\frac{2}{1+\Pi(q^2)}+\xi(q^2)\],\label{linmeq}
\eea 
where we have replaced $M(k^2)$ in the denominator by $m\equiv M(0)$.
In terms of the coordinate space function 
\bea
\psi(r)\equiv \int \frac{{\rm d}^3 p}{(2\pi)^3}\, 
\frac{M(p^2)}{p^2+m^2}\,{\rm e}^{ip\cdot r} \label{fourm},
\eea
the equivalent of Eq.(\ref{linmeq}) is
\bea
\(-\nabla^2+m^2\)\psi(r)=e^2 \int \frac{{\rm d}^3p\,{\rm d}^3k}{(2\pi)^6}
\frac{M(k^2)}{k^2+m^2} \frac{1}{q^2}
\[\frac{2}{1+\Pi(q^2)}+\xi(q^2)\]\,{\rm e}^{ip\cdot r}.\nonu\\
\eea
Performing a shift of integration variable, $p\rightarrow p+k$, 
we find a Schr\"odinger-like equation for the function $\psi(r)$: 
\bea
\(-\nabla^2+V(r)\)\psi(r)=-m^2 \psi(r),\label{schrod1}
\eea
where the potential $V(r)$ is defined as 
\bea
V(r)=-\frac{e^2}{2\pi^2 r}
\int\limits_0^{\infty} {\rm d} p\,
\frac{\sin pr}{p}\[\frac{2}{1+\Pi(p^2)}+\xi(p^2)\]\label{pot}.
\eea
In order to simplify (\ref{pot}), for both the massless and massive case we 
approximate the vacuum polarization by
\bea
\Pi(q^2)=\frac{\alpha}{q+\beta}\label{vpapprox}
\eea 
where $\beta=\sigma m$, $\sigma$ being some fitting constant, and
$\alpha=N_f e^2/8$.
The massless vacuum polarization corresponds to putting $\sigma=0$, while
the value $\sigma=3\pi/4$ corresponds to approximating both infrared and 
ultraviolet behaviors of the exact one-loop expression for 
$\Pi(q^2)$.
The above approximation allows us to perform the integral in $\xi(q^2)$ 
analytically, i.e. with (\ref{vpapprox})
\bea
\int \limits_0^{q^2} \frac{{\rm d}v}{1+\Pi(v)}
=q^2-2\alpha q + 2\alpha(\alpha+\beta)\log \( 1+\frac{q}{\alpha+\beta} \).
\eea
With $\alpha=1$, i.e. $e^2=8/N_f$, the potential can be expressed as
\bea
 V(r)&=&-\frac{8}{\pi^2 N_f r}
\int \limits_0^{\infty}{\rm d} x\,
\frac{\sin xr}{x}\nonu\\
&&\quad\times\,
\Bigg[\frac{2(x+\beta)}{x+1+\beta}
+
\frac{2}{x}-1-\frac{2(1+\beta)}{x^2}\log\( 1+\frac{x}{1+\beta} \)
\Bigg].
\label{vpotgc1}
\eea
The last expression can be rewritten in a more convenient form: 
\bea
&&V(r)=-\frac{4}{\pi N_f r}\frac{\beta}{\beta+1}-\frac{16}{\pi^2 N_f r}
\int \limits_0^{\infty}{\rm d} x\sin xr\nonu\\
&&\times\,\Bigg[\frac{1}{(1+\beta)(x+1+\beta)}+\frac{1}{x^2}
-\frac{1}{2(1+\beta)x}
-\frac{(1+\beta)}{x^3}\log\( 1+\frac{x}{1+\beta} \)
\Bigg].\label{vpotgc2}
\eea
We can get rid of the oscillatory behavior of the sine 
by performing a contour rotation in the 
lower right quadrant 
of the complex-plane of $x$. 
So
\bea
V(r)&&=-\frac{4}{\pi N_f r} \frac{\beta}{1+\beta}
-\frac{16}{\pi^2 N_f r}\int \limits_0^\infty {\rm d} y\,
\exp(-yr)\nonu\\
&&\quad\times\,\[\frac{1}{y^2+(1+\beta)^2}-\frac{1}{y^2}+\frac{(1+\beta)}{y^3}
\arctan \frac{y}{1+\beta}\].\label{potential}
\eea
Let us consider first the case of vacuum polarization by massless
fermions, i.e. $m=0$, which coincides precisely with the approximation
used by Appelquist, Nash and Wijewardhana \cite{appel2} in momentum space. 
The potential (\ref{potential}) takes the form
\bea
V(r)=-\frac{16}{\pi^2N_fr}\int\limits_0^\infty dy\exp(-yr)\biggl[\frac
{1}{y^2+1}-\frac{1}{y^2}+\frac{1}{y^3}\arctan y\biggr].
\label{mslesspot}
\eea
This potential behaves at large distances as
\bea
V(r)\sim -\frac{\lambda_1}{r^2}, \quad \lambda_1=\frac{32}{3\pi^2N_f},
\eea
whereas at small distances it has Coulomb-like behavior:
\bea
V(r)\sim -\frac{\lambda_2}{r},\quad \lambda_2=\frac{4}{\pi N_f}.
\eea
 From the theory of the Schr\"odinger equation for 
such kinds of potential, it is known \cite{Landau,Perel} that there is 
an infinite number of bound states if $\lambda_1>1/4$ ($N_f<N_c=128/3\pi^2$). 
However a finite number of bound states might exist for $\lambda_1<1/4$ as 
well. Let us show that for the potential under consideration there are no 
bound states at all for $\lambda_1<1/4$. Indeed, one can prove the following 
inequality
\bea
V(r) > -\frac{\lambda_1}{r}\int\limits_0^{\infty}{\rm d}y\,
\frac{\exp(- y r )}{y^2+1}>-\frac{\lambda_1}{r^2}. 
\label{inelam}
\eea
Since the Schr\"odinger equation
\bea
\(-\nabla^2-\frac{\lambda_1}{r^2}\)\psi(r)=-m^2\psi(r) \label{scheq}
\eea 
has no bound states for $\lambda_1<\frac{1}{4}$ \cite{Landau}, it is 
evident from the inequality (\ref{inelam}) that the Schr\"odinger equation 
(\ref{schrod1}) with the potential $V(r)$ (\ref{mslesspot}) has none either. 
Thus we have proved that the critical number of fermions,
\bea
N_c=\frac{128}{3\pi^2}\approx 4.32304,\label{ncrit}
\eea
does indeed correspond to the appearance of an infinite number of bound states 
in an effective Schr\"odinger equation in $r$-space, if we consider the vacuum
polarization by massless fermions. 
The critical number of fermions (\ref{ncrit}) 
coincides with the result 
obtained by Nash in a different way \cite{Nash2} and that was 
claimed as being gauge invariant to leading order in the $1/N_f$ expansion.

The quasi-classical approach gives the
following spectrum of bound states for potentials behaving at large
distances as $-\lambda/r^2$ \cite{Perel}:
\bea
E_n=-m^2_n\sim-\exp\biggl[-\frac{2\pi n}{\sqrt{\lambda_1-1/4}}\biggr]
\eea 
at $n>>1$, coinciding in fact with the result of Refs.\cite{appel2,Nash2}. 
As was 
pointed out in \cite{appel2}, this spectrum is closely connected with 
approximate scale invariance of 
${\rm QED_3}$ at intermediate distances 
$1<r<1/m$ \footnote{We note that in ${\rm QED_4}$ it is 
exactly the scale invariance
of the ladder and/or quenched approximation which is responsible for 
exponential-like behavior of a dynamical fermion mass. For physical 
explanation of this fact see Ref.\cite{FGMS}. Further discussion of the 
role of scale invariance in ${\rm QED_4}$ can be found in Refs.
\cite{Bardeen,Holdom,AtkGusMar}.}.  
However 
the quasi-classical theory does not tell us anything about lower levels of the 
Schr\"odinger equation. It is clear, also, that the proof given above breaks 
down in the case of massive fermion vacuum polarization, since 
the potential (\ref{potential}) 
has a Coulomb-like tail at large distances. The latter is simply 
a reflection of the fact that including a dressed fermion propagator restores 
the confinement property of ${\rm QED_3}$ 
in the sense that a logarithmic term reappears 
in the real potential at large $r$ \cite{Burden}. In the next section we 
shall solve the Schr\"odinger equation with the potential (\ref{potential}) 
and study the scaling properties of the dynamical mass near $N_c$.

\section{Scaling Properties near Criticality}

In this section we study the scaling properties of the dynamical mass
near $N_c$, for both massless and massive vacuum polarization.
For large $r$, $V(r)$ (\ref{potential}) has the following form
\bea
V(r)\approx-\frac{\lambda_1}{r^2}-\frac{\lambda_3 m}{r},
\qquad r\gg {\cal O}(1),
\eea
where 
\bea
\lambda_3=\frac{4 \sigma }{\pi N_f},\qquad \beta=\sigma m.
\eea
In deriving this we neglected terms $\beta$ with respect to $1$, since
we consider $V(r)$ near $N_c$, where $m$ is very small (and so is $\beta$).
For large $r$ we have thus a $-1/r^2$ potential with a small Coulomb
potential which starts to contribute for 
$r\geq 1/\beta$.
For small $r$, $V(r)$ behaves like a Coulomb potential plus a constant part
\bea
V(r)\approx -\frac{\lambda_2}{r}+\frac{3}{2}\lambda_1,
\qquad r\ll {\cal O}(1).
\eea
We split our problem into two regions, for large $r$, and small $r$.
Neglecting angular dependence, which is not relevant for our purpose, 
we obtain 
\bsea
\frac{{\rm d}^2\psi}{{\rm d} r^2}
+\frac{2}{r}\frac{{\rm d}\psi}{{\rm d}r}
+\(\frac{\lambda_1}{r^2}+\frac{\lambda_3 m}{r}-m^2\)\psi&=&0, \qquad r \gg 1,
\slabel{large}\\
\frac{{\rm d}^2\psi}{{\rm d} r^2}
+\frac{2}{r}\frac{{\rm d}\psi}{{\rm d}r}
+\(\frac{\lambda_2}{r}-\tilde m^2 \)\psi&=&0,\qquad r\ll 1,
\slabel{small}
\esea
where $\tilde m^2=\frac{3}{2}\lambda_1+m^2$. 
With $\psi(r)=f(r)/r$, the equations can be expressed
as Whittaker equations
\bsea
\frac{{\rm d}^2 f }{{\rm d} x^2} 
+\( \frac{\lambda_1}{x^2 }+\frac{\lambda_3}{2x}-\frac{1}{4}\) f=0, &&
\quad x=2 m r,\quad r\gg 1,\label{largef}\\
\frac{{\rm d}^2 f }{{\rm d} x^2} 
+\( \frac{\lambda_2}{2 \tilde m x }-\frac{1}{4}\) f=0, &&
\quad x=2\tilde m r,\quad r\ll 1, \label{smallf}
\esea
where the
general form of a Whittaker differential equation is defined as \cite
{Erdelyi}
\bea
\frac{{\rm d}^2 f }{{\rm d} x^2} 
+\(\frac{\frac{1}{4}-\mu^2}{x^2}+\frac{\kappa}{x}-\frac{1}{4}\) f=0.
\label{whit}
\eea
Two independent solutions of (\ref{whit}) are the Whittaker functions
\bsea
M_{\kappa,\mu}(x)&=&{\rm e}^{-\frac{x}{2}} x^{\frac{c}{2}} \Phi(a,c;x),
\slabel{whitm}\\
W_{\kappa,\mu}(x)&=&{\rm e}^{-\frac{x}{2}} x^{\frac{c}{2}} \Psi(a,c;x)
\slabel{whitw},
\esea
where
\bsea
\kappa&=&\frac{c}{2}-a,\\
\mu&=&\frac{c}{2}-\frac{1}{2},
\esea
$\Psi$ and $\Phi$ are confluent hypergeometric functions.
The solution of Eq.(\ref{large}) 
that is regular at $r=\infty$, has the form 
\bea
\psi_1(r)={\rm e}^{-mr} r^{-\frac{1}{2}+i\nu}
\Psi \(\frac{1}{2}-\frac{\lambda_3}{2}+i\nu,1+2i\nu;2mr\),\quad r \gg 1,
\label{sollarge}
\eea
where $\nu=\sqrt{\lambda_1-\frac{1}{4}}\,,$ which vanishes when
$N_f\rightarrow N_c$.
Eq.(\ref{small}) has a solution of the type (\ref{whitm}), which is 
regular at $r=0$:
\bea
\psi_2(r)={\rm e}^{-\tilde m r} 
\Phi \(1-\frac{\lambda_2}{2\tilde m}, 2 ;2\tilde m r\),\quad r \ll 1.
\label{solsmall}
\eea
Since these two different solutions represent 
one solution of our original problem (\ref{schrod1}), 
we have to match them at some intermediate point
$r=r_0={\cal O}(1)$. A choice of $r_0$ corresponding to Appelquist et al.
would be $r_0=1/\alpha=1$.
But for the time being we keep an arbitrary matching point.
As matching condition we take 
\bea
\frac{{\rm d}}{{\rm d} r}\[\log  \frac{\psi_2(r)}{ \psi_1(r)}\]_{r=r_0}=0.
\label{match}
\eea
With
\bea
&&\Phi^{\prime}(a,c;x)=\frac{a}{c}\Phi(a+1,c+1;x),\\
&&\Psi^{\prime}(a,c;x)=-a\Psi(a+1,c+1;x),
\eea
the matching condition (\ref{match}) reads
\bea
&&2m\(\frac{1}{2}-\frac{\lambda_3}{2}+i\nu\)
\frac{\Psi   \( \frac{3}{2}-\frac{\lambda_3}{2}+i\nu, 2+2i\nu; 2mr_0\)}{
\Psi\(\frac{1}{2}-\frac{\lambda_3}{2}+i\nu, 1+2i\nu; 2mr_0\)}
=\nonu\\
&&=-\frac{1}{r_0}\(\frac{1}{2}-i\nu\)-m+\tilde m-
\(\tilde m- \frac{\lambda_2}{2}\)\frac{\Phi\(2-\frac{\lambda_2}{2\tilde m},
3; 2\tilde mr_0\)}{
\Phi\(1-\frac{\lambda_2}{2\tilde m},2; 2\tilde m r_0\)}.
\eea
Since we consider small $m$, the dependence on $m$ on right-hand side 
of the previous equation vanishes.
We define 
\bea
d=\tilde m-
\(\tilde m- \frac{\lambda_2}{2}\)\frac{\Phi\(2-\frac{\lambda_2}{2\tilde m},
3; 2\tilde mr_0\)}{
\Phi\(1-\frac{\lambda_2}{2\tilde m},2; 2\tilde m r_0\)},
\eea
with $\tilde m=\sqrt{\frac{3\lambda_1}{2}}$. 
Now $2mr_0\ll 1$, so for the $\Psi$-function we can use the asymptotic 
form for small argument, i.e.
\bea
\Psi(a,c;x)\approx_{_{_{_{\hspace{-4mm}{{x\rightarrow}
{ 0}}}}}}
\frac{\Gamma(1-c)}{\Gamma(a-c+1)}+\frac{\Gamma(c-1)}{\Gamma(a)}x^{1-c}.
\eea
The matching condition then reads
\bea
&&\hspace{-5mm}\(\frac{1}{2} -\frac{\lambda_3}{2}+i\nu\)
\[\frac{\Gamma\(-1-2i\nu\)}{\Gamma\(\frac{1}{2}-\frac{\lambda_3}{2}-i\nu\)}
(2mr_0)^{1+i\nu}+
\frac{\Gamma\(1+2i\nu\)}{\Gamma\(\frac{3}{2}-\frac{\lambda_3}{2}+i\nu\)}
(2mr_0)^{-i\nu}
\] =\nonu\\
&&\hspace{-1cm}=\(dr_0-\frac{1}{2}+i\nu\)\[\frac{\Gamma\(-2i\nu\)}{\Gamma
\(\frac{1}{2}-\frac{\lambda_3}{2}-i\nu\)}(2mr_0)^{i\nu}+
\frac{\Gamma\(2i\nu\)}{\Gamma\(\frac{1}{2}-\frac{\lambda_3}{2}+i\nu\)}
(2mr_0)^{-i\nu}\],
\eea
which can be rewritten as 
\bea
&&\hspace{-3mm}\(\frac{1}{2}+i\nu\)\frac{\Gamma\(2i\nu\)}{
\Gamma\(\frac{1}{2}-\frac{\lambda_3}{2}+i\nu\)}(2mr_0)^{-i\nu}+
\(\frac{1}{2}-i\nu\)\frac{\Gamma\(-2i\nu\)}{
\Gamma\(\frac{1}{2}-\frac{\lambda_3}{2}-i\nu\)}(2mr_0)^{i\nu}=\nonu\\
&&\hspace{-5mm}=dr_0\[\frac{\Gamma\(2i\nu\)}{
\Gamma\(\frac{1}{2}-\frac{\lambda_3}{2}+i\nu\)}(2mr_0)^{-i\nu}+
\frac{\Gamma\(-2i\nu\)}{
\Gamma\(\frac{1}{2}-\frac{\lambda_3}{2}-i\nu\)}(2mr_0)^{i\nu}\].
\label{matcon}
\eea
With the use of
\bea
a {\rm e}^{i\phi}+\bar a {\rm e}^{-i\phi}=2|a|\cos(\phi+\theta),\quad 
\theta = \arg(a),
\eea
the equation (\ref{matcon}) can be reduced to the following one
\bea
\tan\(\nu\log\frac{1}{2mr_0}+\Sigma(\nu)\)=\frac{2\nu}{2dr_0-1},
\label{tana}
\eea
where
\bea
\Sigma(\nu)=\arg\(\frac{\Gamma(1+2i\nu)}{\Gamma({1\over2}
-{\lambda_3\over2}+i\nu)}\).
\eea
Eq.(\ref{tana}) then gives for $m$: 
\bea
 m =\exp\biggl[-\frac{n\pi}{\nu}-\log(2r_0) + {1\over\nu}\arctan
{\frac{2\nu}{1-2dr_0}}+{1\over\nu}\Sigma(\nu)\biggr],
\label{masseq}
\eea
with $n$ a positive integer.
In the limit $\nu\to 0$, Eq.(\ref{masseq}) takes the form
\bea
m=\exp\(-\frac{2\pi n}{\sqrt{\frac{N_c}{N_f}-1}}+b\),\label{scalelaw}
\eea
where
\bea
b=-\log(2r_0) + \frac{2}{1-2dr_0}
-2\gamma+\psi\(\frac{1}{2}
-\frac{\lambda_3}{2}\),
\eea
with $\psi$ the di-gamma function, $\psi(z)=\Gamma^{\prime}(z)/\Gamma(z)$, 
and where $\gamma$ is the Euler constant.
Only the solution with the largest value of $m<1$ ($n=1$) corresponds to the
ground state, since it has the lowest energy.

With the help of Mathematica we can calculate 
$b$ explicitly for various values of the parameter $\sigma$ ($r_0=1$).
With $\nu=0$
\bea
&&\lambda_1=\frac{1}{4},\qquad
\lambda_2=\frac{3\pi}{32},\qquad
\lambda_3=\frac{3\pi\sigma}{32},\qquad
\tilde m=\sqrt{\frac{3}{8}},
\eea
this gives
\bsea
\mbox{massless fermion loops:}\quad &&\sigma=0,\qquad b=2.214,\nonu\\ 
\mbox{massive fermion loops:}\quad &&\sigma=\frac{3\pi}{4},\qquad b=7.136.
\esea

Thus we have found that for a vacuum polarization with massive fermions, 
the infrared dynamical mass $m$ obeys an exponential scaling law
near the critical point $N_c$ despite the presence of a Coulomb tail in
the potential (\ref{potential}). This means that chiral symmetry breaking
takes place essentially at intermediate distances $1<r<1/m$. 
For the massless vacuum polarization, this was first pointed out
by Appelquist et al. \cite{appel2}.
The main difference between a vacuum polarization with massive (parameter
$\sigma\neq0$) or with massless fermions ($\sigma=0$) lies in the scaling 
factor $b$, which is larger in the case of massive fermion loops in the 
vacuum 
polarization.
We note here that our real space approach automatically takes into account the 
ultra violet tail of the integral equation in momentum space to which
an attention was paid in Ref.\cite{KondoMaris}. Moreover,
in comparison with their massless vacuum polarization calculations, our 
analytical solution gives the scaling factor $b$ closer to exact 
computation of that from whole nonlinear equation (compare our $b=2.214$
with Kondo and Maris' $b=3.94$ \cite{KondoMaris} 
against exact $b=1.842$).

\section{Numerical Calculations in Momentum\\ Space}

In this section we discuss the numerical calculation
of the scaling law of the dynamical mass near criticality 
for
a vacuum polarization with respectively 
massive and massless fermion loops.
The nonlinear DSE can be written in the following way, with $\alpha=1$
\bea
M(p^2)=\frac{2}{\pi^2 N_f} 
\int\limits_0^{\infty} {\rm d}k\, 
\frac{k^2 M(k^2)}{k^2+M^2(k^2)} K(p^2,k^2,m),\label{mktheta}
\eea
where the kernel $K$ is given by
\bea
K(p^2,k^2,m)=  \int\limits_0^{\pi} {\rm d}\theta\,
\sin\theta \frac{4}{q^2}
\[\frac{1}{1+\Pi(q^2)}-\frac{1}{2}\int_0^1\frac{{\rm d} y}{1+\Pi(q^2y) }\].
 \label{kernel}
\eea
with $q^2=k^2+p^2-2pk \cos\theta$, and
we recall the expression for the one-loop vacuum polarization, 
\bea
\Pi(q^2)=\frac{ 2 }{\pi q^2}\[2 m
+ \frac{q^2-4m^2}{q}\arctan\(\frac{q}{2m}\)\].
\eea
In the case of massless fermions in the vacuum polarization, 
the $y$-integration in (\ref{kernel}) can be performed explicitly.
It reduces to
\bea
K(p^2,k^2,0)=
\int\limits_0^{\pi} {\rm d}\theta\,
\sin\theta \frac{4}{q^2}
\[\frac{q}{1+q}-\frac{1}{2}+\frac{1}{q}-\frac{1}{q^2}\log\(1+q\)\],
\label{kernelml}
\eea
which does not depend on the dynamical mass, so that it only 
has to be calculated once.
It is convenient to perform the calculations 
on a $\log p^2$ scale in order to see more details of the structure.
\bea
M(x)=\frac{1}{\pi^2 N_f} 
\int\limits_{\log\mu^2}^{\log\Lambda^2} {\rm d}t\, 
\frac{y\sqrt{y} M(y)}{y+M^2(y)} K(x,y,m),\label{mktheta2}
\eea
where $x=p^2$, and $y=k^2=\exp(t)$.
We have introduced  infrared and ultraviolet cutoffs, $\mu$ and 
$\Lambda$, respectively, which should satisfy 
$\mu\ll M(0)$, and $\Lambda\gg \alpha=1$.

The basic idea of our numerical calculations is that
we solve Eq.(\ref{mktheta}) iteratively as an integral equation.
First we calculate the kernel $K$ for all grid points $x$ and $y$ 
with some initial value for $m$.
Then we iterate the integral equation 
until it has reached some convergence criterion.
Then we recalculate the kernel with the new $m=M(0)$, and iterate again.
This procedure we repeat to  the point were the complete system 
has converged.
To obtain a satisfactory numerical procedure, the integrals in $K$
are performed
using Gauss-Legendre quadrature, which works well since our 
integrands in (\ref{mktheta2})  are sufficiently smooth.  
For $M$ at the grid points $x$ we use  
cubic spline interpolation on a $\log x$-scale.
The $t$-integration is performed using Gauss-Legendre quadrature between
the grid points $x$,
\bea
M(x_i)=\frac{1}{\pi^2 N_f} \sum_{j=1}^{n-1}
\int\limits_{\log x_j}^{\log x_{j+1}} {\rm d}t\,
 \[\frac{y\sqrt{y} M(y)}{y+M^2(y)} K(x_i,y,m)\],\quad i=1,\dots,n,
\eea
where $x_1=\mu^2$, and $x_n=\Lambda^2$.
With this procedure we find that for calculation near the critical point
$N_c$, we need at least $128$ grid points for $x$.
We have calculated the mass function $M(x)$ for a series 
of values of $N_f$ near $N_c$.
For all cases it appears that
the infrared $m$ obeys the exponential scaling law.
This has been verified using a least squares fit for the numerical data.
We have fitted
the infrared mass $m$, as a function of $N_f$, with the following function 
\bea
\log m(N_f)=-\frac{2\pi a}{\(c/N_f-1\)^d}+b\,,\label{expbeh}
\eea
where $a$, $b$, $c$, and $d$ are the parameters to be fitted.
For the nonlinear massless case the least squares fitting 
of the data gives 
\bea
a=0.9907,\qquad
b=1.720,\qquad
c=4.32312,\qquad
d=0.5018,
\eea
and with massive fermion loops
\bea
a=1.055,\qquad
b=3.494,\qquad
c=4.32315,\qquad
d=0.4893,
\eea
which is shown in Fig.~\ref{fig2}.
This fitting proves the exponential scaling law (\ref{scalelaw}).
To obtain values of $b$ and compare them with the analytic calculation 
we take $a=1$, $c=N_c=128/3\pi^2$, $d=1/2$.
The numerical results are:
\bea
\mbox{massless fermion loops:}\quad&& b=1.842,\nonu\\ 
\mbox{massive fermion loops:} \quad&& b=2.955.
\eea
We also investigated the difference between a linearized approximation 
(\ref{linmeq}) and the nonlinear DSE (\ref{meqeucl}). 
Qualitively the linearized approximation is equivalent to the more 
realistic nonlinear equation.
Quantitively the scaling factor $b$ 
is a few percent smaller for the linearized model.

In Fig.\ref{fig3} we have plotted the dynamical fermion mass function 
$M(p^2)$, computed for various values of the fermion number $N_f$.
Apart of $N_c$ it has behavior reminiscent to that in the quenched 
approximation \cite{Hosh,Kondo}: the mass function is constant at momenta  
$p\ll \alpha$ and decreases as $\alpha^3/p^2$ at large momenta.
The mass function changes its behavior at momentum $p\simeq\alpha$,
a natural scale of the model.
Close to critical number of fermions, 
$N_f\alt N_c$,
we observe three different regions in behavior of 
$M(p^2)$:
$M(p^2)\simeq m$ at $p\leq m$, $M(p^2)\sim m^{3/2}/\sqrt{p}$ at 
$m\leq p\leq \alpha$ and $M(p^2)\sim m^3/p^2$ at $p > \alpha$.
The scale $m$ is exponentially small in comparison with the intrinsic 
scale $\alpha$, see Eq.(\ref{expbeh}).
Fig.\ref{fig3} shows convincingly how a hierarchy of scales 
can occur in a theory under specific conditions (in our case 
when $N_f$ is close to $N_c$).
This might be important for further understanding a hierarchy 
of scales in unified theories.

\section{Conclusion and Discussion}

  In the present paper we investigated $\chi{\rm SB}$ in
(2+1)-dimensional $\rm QED$ with $N_f$ number of four-component Dirac
fermions. We have pointed out one possible scheme for truncating an
infinite chain of SDE, which is consistent with WT identity and which
seems to have been overlooked so far. Further simplification of that coupled
system of SDE led to studying the SD equation for a fermion mass
function where the full vacuum polarization with dynamically massive 
fermions has been taken into account. The latter proves to be important
for keeping the confinement property of the model at large distances.
The study of the linearized equation for a fermion mass function has
been performed in  real space,and  that allowed us to avoid making 
drastic approximations in order to get analytical results. Our
analytical solution fits nicely numerical calculations of the full nonlinear
integral equation for the mass function.

Further investigation of aforementioned truncated system of SDE is now
in progress.

\section{Acknowledgements}
We would like to express sincere thanks to David Atkinson for 
many stimulating discussions and for correcting the manuscript.
We thank Y. Hoshino, who took part in investigations at 
early stages of this work,  for 
collaboration.
We are grateful also to Pieter Maris and Volodya Miransky for valuable remarks.
V.P.G. is grateful to the members of the Institute for Theoretical Physics
of the University of Groningen for their hospitality.
He wishes to acknowledge the Stichting FOM (Fundamental Onderzoek der 
Materie), financially supported by the Nederlandse Organisatie voor 
Wetenschappelijk Onderzoek, for its support.
The work of one of us (V.P.G.) was supported in part by Grant INTAS-93-2058
``East-West network in constrained dynamical systems''.

\newpage

\begin{figure}
\caption{Truncated system of Schwinger-Dyson equations.}\label{fig1}
\bigskip
\caption{The numerical results for $M(0)$ as function of $N_f$ for 
massive and massless fermion loops in vacuum polarization.}\label{fig2}
\bigskip
\caption{Dynamical mass function $M(p^2)$ for $N_f=1.0$, 
$2.0$, $3.0$, $4.0$.}
\label{fig3}
\end{figure}

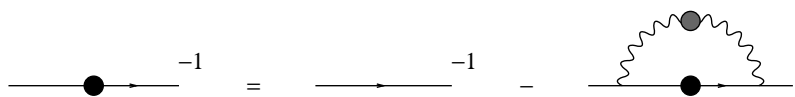
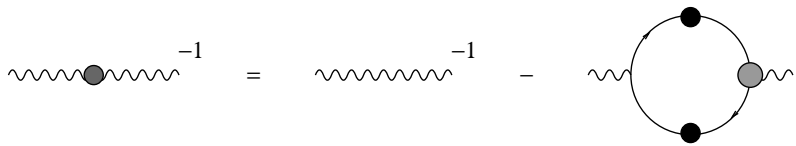
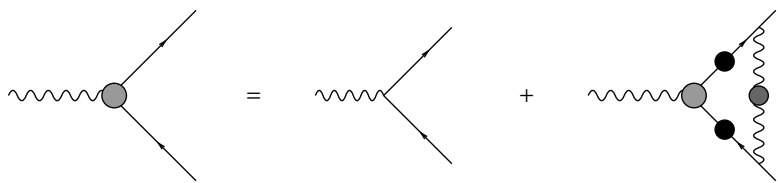
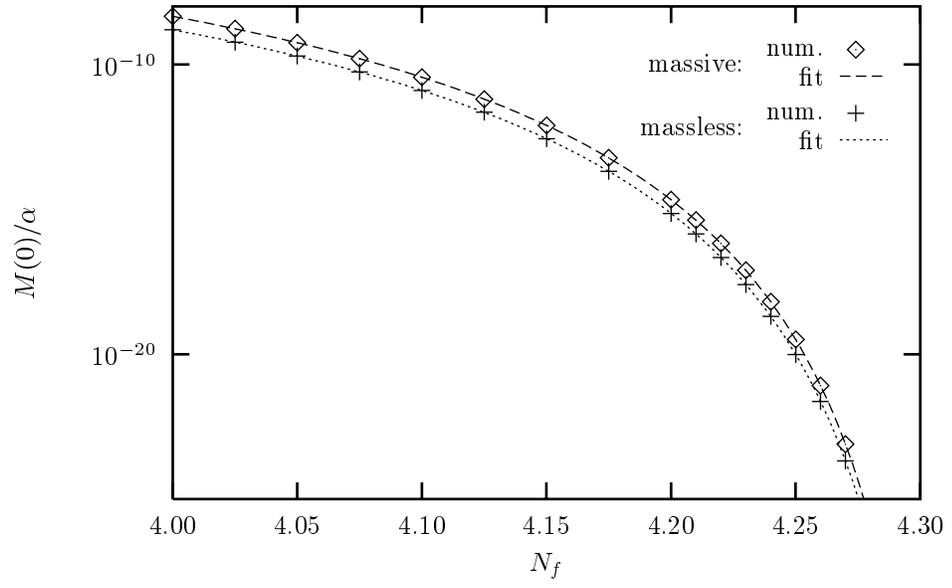
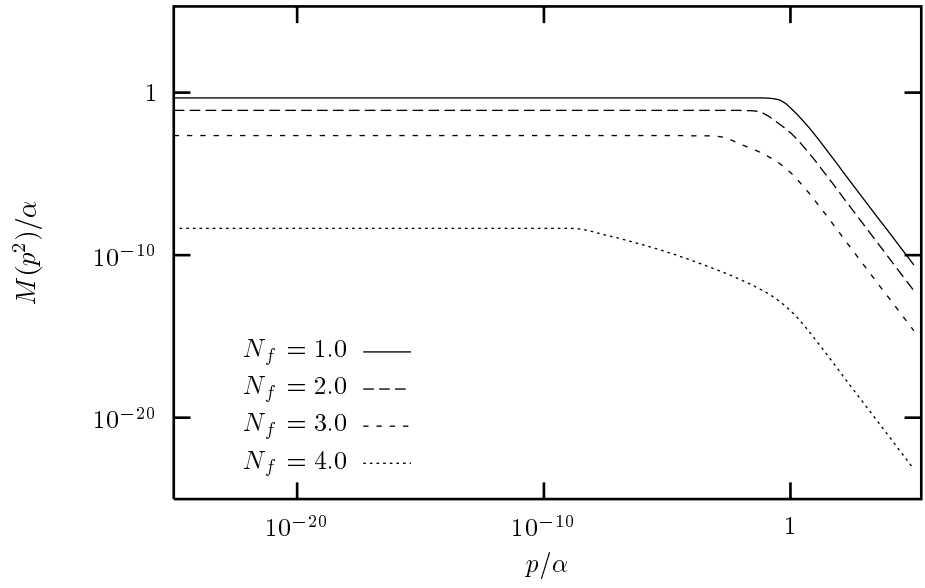
\end{document}